\documentclass[conference]{IEEEtran}
\usepackage{cite}
\usepackage{amsmath,amssymb,amsfonts}
\usepackage{graphicx}
\usepackage{textcomp}
\usepackage{tabularx,booktabs}
\usepackage{makecell}
\usepackage{braket}
\usepackage{xurl}
\usepackage{algorithm}
\usepackage{algpseudocode}
\usepackage[caption=false,font=footnotesize]{subfig}
\newcommand{\PARstart}[2]{#1#2}
\makeatletter
\newcommand{\finalfigurecaption}{\def\@captype{figure}\caption}
\makeatother

\begin{document}
\pagestyle{empty}
\title{An Adaptive Quantum Search Framework under Unknown Solution-density: Minimizing Expected Queries to the First Solution
}

\author{
\IEEEauthorblockN{Wiwat Asawalertsak}
\IEEEauthorblockA{\textit{Quantum Computing and Information Research Centre}\\
\textit{King Mongkut's University of Technology Thonburi}\\
Bangkok, Thailand\\
wiwat.asawa@mail.kmutt.ac.th}
\and
\IEEEauthorblockN{Rajchawit Sarochawikasit}
\IEEEauthorblockA{\textit{Department of Computer Engineering}\\
\textit{Quantum Computing and Information Research Centre}\\
\textit{King Mongkut's University of Technology Thonburi}\\
Bangkok, Thailand\\
rajchawit.sar@mail.kmutt.ac.th\\
(Corresponding author)}
}
\maketitle
\thispagestyle{empty}

\begin{abstract}
    Grover's algorithm provides a mathematically optimal quadratic speedup for unstructured search, yet its standard formulation relies on a critical prerequisite: prior knowledge of the target solution count to determine the number of algorithm iterations. Although numerous strategies have been developed to identify or bypass this parameter, comparing them directly is nontrivial. This study investigates and computationally optimizes 3 prominent frameworks: Boyer-Brassard-H{\o}yer-Tapp, Fixed-Point search, and Density Estimation Methods. Density-specific benchmarks select the best tested method and configuration to minimize expected first-solution queries under 2 initialization-cost conventions. To translate these theoretical benchmarks into practical utility, a hybrid quantum-classical adaptive framework integrating Hill-climbing-assisted Grover Adaptive Search with state-space density estimation modeled via the Wang-Landau algorithm is proposed as a potential use-case. This implementation operates directly on the optimized query methods established earlier. Analytical models support the approach; ideal simulations show 18.8\%--63.4\% lower mean accumulated query costs than the tested randomized and fixed-point baselines. 
\end{abstract}


\section{Introduction}
    \PARstart{S}{earching} is one of the most basic computational problems. Many real-world computationally expensive tasks can be mapped into a kind of searching, including optimization problems such as the Traveling Salesman Problem or the Knapsack Problem~\cite{garey1979}. In the context of computer science, searching an unstructured database of $N$ entries with 1 target requires $\Theta(N)$ queries, and no classical approach can asymptotically surpass this bound~\cite{nielsen2010}. However, this limitation is relaxed when it comes to quantum computing. Thanks to the abilities of quantum computing that allow a single oracle call to act on a superposition of all $N$ index states, together with phase interference and amplitude amplification, quantum algorithms such as Grover Search are proved mathematically to achieve $O(\sqrt{N})$ when executed on noiseless quantum hardware~\cite{grover1996,brassard2002}.

    Expected-query optimization has been studied beyond asymptotic bounds. Boyer \textit{et al.}~\cite{boyer1998} analyzed how the Grover iteration count can be chosen to minimize the expected number of iterations when unsuccessful runs are restarted. For fixed-point search, Yoder \textit{et al.}~\cite{yoder2014} established success guarantees with optimal query scaling for a prescribed error tolerance, whereas Nagy \textit{et al.}~\cite{nagy2024} numerically optimized expected search cost in the sparse-density limit and developed an increasing-depth strategy. Giuffrida \textit{et al.}~\cite{giuffrida2022} studied strategies for adapting Grover iteration counts and stopping policies in Grover adaptive search (GAS) based on algorithm evolution and previous outcomes. 

    This study formulates parameter optimization for the Boyer--Brassard--H{\o}yer--Tapp (BBHT) method~\cite{boyer1998}, fixed-point search (FP)~\cite{grover2005,yoder2014}, and density estimation methods (DEM)~\cite{brassard2002,suzuki2020,grinko2021}. Here, the contribution is a density-specific expected-cost comparison across these methods over a common density range, with state preparation either excluded from the cost or charged 1 additional unit per trial. The resulting benchmark tables identify the preferred method and configuration at each evaluated density and provide a potential use-case example of adaptive selection in GAS assisted by hill climbing (HC) and Wang--Landau (WL) density estimation. The effect of these selections on the accumulated query cost is evaluated using analytical cumulative-query models and ideal simulations, separately from the per-density benchmarks.

\section{Background Knowledge and Related Works}
\subsection{Standard Grover Algorithm}
    In generalized quantum search, the quantum state can be visualized as a vector rotating within a 2-dimensional real plane. This plane is spanned by 2 orthonormal basis vectors: the uniform superposition of all unmarked states, and the uniform superposition of all marked states~\cite{grover1996,brassard2002}. For an $n$-qubit register with $M$ marked states,
    \begin{equation}
\begin{aligned}
        N&=2^n,\quad 1\leq M<N,\quad a=M/N,\quad\theta=\arcsin\sqrt a,\\
        |U\rangle&=\frac{1}{\sqrt{N-M}}\sum_{x\notin\text{targets}}|x\rangle,
        \quad |T\rangle=\frac{1}{\sqrt M}\sum_{x\in\text{targets}}|x\rangle,\\
        |\psi_0\rangle&=H^{\otimes n}|0\rangle^{\otimes n}
        =\cos\theta\,|U\rangle+\sin\theta\,|T\rangle.
    \end{aligned}
\label{eq:grover_basis}
\end{equation}
    The Grover iteration operator can be generalized as $G(\beta, \alpha)$ using parameterized phase shifts $\beta$ and $\alpha$~\cite{hoyer2000,long2001}:
    \begin{equation}
\begin{aligned}
        O(\beta)&=I+(e^{i\beta}-1)|T\rangle\langle T|,\\
        D(\alpha)&=I+(e^{i\alpha}-1)|\psi_0\rangle\langle\psi_0|,\\
        G(\beta,\alpha)&=D(\alpha)\cdot O(\beta).
    \end{aligned}
\label{eq:generalized_grover}
\end{equation}
    Standard Grover's algorithm is the case in which the phase parameters are fixed at $\beta = \pi$ and $\alpha = \pi$.
    It is therefore convenient to define the Grover rotation operator $G\equiv -G(\pi,\pi)$, so that $G^{k} = (-1)^{k} G(\pi,\pi)^{k}$; the operators $G^k$ and $G(\pi,\pi)^k$ differ only by a global phase and are physically indistinguishable under measurement.
    \begin{equation}
\begin{aligned}
        |\psi_k\rangle&=G^k|\psi_0\rangle
        =\cos((2k+1)\theta)|U\rangle\!+\sin((2k+1)\theta)|T\rangle,\\
        P(k)&=|\langle T|\psi_k\rangle|^2=\sin^2((2k+1)\theta),\\
        k^*&=\left\lfloor\frac{\pi}{4\theta}-\frac12\right\rceil.
    \end{aligned}
\label{eq:standard_grover}
\end{equation}

    Owing to the periodicity of $\sin^{2}$, iterating beyond $k^{*}$ rotates the state past $|T\rangle$ and initially degrades the success probability, which is precisely why prior knowledge of $M$ is required. For small target densities ($M \ll N$), the small-angle approximation $\theta \approx \sin\theta = \sqrt{M/N}$ applies, yielding at leading order $k^*\approx(\pi/4)\sqrt{N/M}\in O(\sqrt{N/M})$, which reduces to the familiar $O(\sqrt{N})$ query count when $M$ is constant. This scaling is optimal, as no quantum algorithm can solve unstructured search with fewer than $\Omega(\sqrt{N})$ oracle queries~\cite{bennett1997,zalka1999}.

\subsection{Methods for Unknown Solution Density}

    In the standard Grover scheme, fixing the phase parameters at $\beta = \pi$ and $\alpha = \pi$ requires exact knowledge of the number of target solutions $M$ in order to calculate the iteration count $k^{*}$. Since $M$ is unavailable in many practical scenarios, a number of subsequent techniques have been developed to address this limitation. 

    \subsubsection{The Boyer--Brassard--H{\o}yer--Tapp Method}

        To overcome this limitation, Boyer \textit{et al.} introduced a randomized, adaptive framework~\cite{boyer1998}. Instead of aiming for a fixed target angle, the algorithm chooses the number of Grover iterations stochastically from a dynamically expanding search bound. Let $\lambda$ be a constant scaling factor; the procedure is structured as shown in Algorithm~\ref{alg:bbht}.

        \begin{algorithm}[t]
            \caption{BBHT Algorithm}
            \label{alg:bbht}
            \begin{algorithmic}[1]
                \State Set the initial maximum iteration bound $m = 1$.
                \Loop
                    \State Choose an integer $k$ uniformly at random from the range $k \in \{0, 1, 2, \dots, \lceil m \rceil - 1\}$
                    \State Apply $k$ standard Grover iterations $G^{k}$ to the uniform superposition $|\psi_0\rangle$
                    \State Perform a measurement on the register in the computational basis to obtain an outcome $x$
                    \If{$x$ is a valid target solution}
                        \State Output $x$ and terminate.
                    \Else
                        \State Update the maximum bound geometrically: $m = \min(\lambda m, \sqrt{N})$
                    \EndIf
                \EndLoop
            \end{algorithmic}
        \end{algorithm}

        Note that the bound $m$ is real-valued and is capped at $\sqrt{N}$, which prevents the sampling range from growing without limit once it is large enough to accommodate the smallest admissible target count $M = 1$. In the degenerate case $M = 0$ the loop never terminates, so a practical implementation must impose an external cutoff on the total number of queries~\cite{boyer1998}.

        Because the exact number of iterations $k$ is drawn uniformly at random, the success probability of any given trial is the average success probability over the sampling range $\{0, 1, \dots, m-1\}$. For a fixed integer bound $m$, the average success probability $P_{\text{succ}}(m)$ is therefore the uniform mean of the individual iteration success probabilities:

        \begin{equation}
P_{\text{succ}}(m) = \frac{1}{m} \sum_{k=0}^{m-1} \sin^{2}\bigl((2k+1)\theta\bigr).
\label{eq:bbht_average_probability}
\end{equation}

        By applying trigonometric product-to-sum identities, this discrete summation evaluates to the closed-form expression

        \begin{equation}
P_{\text{succ}}(m) = \frac{1}{2} - \frac{\sin(4m\theta)}{4m\sin(2\theta)},
\label{eq:bbht_closed_probability}
\end{equation}

        \noindent which holds for any $\theta$ with $\sin(2\theta) \neq 0$~\cite{boyer1998}. This expression motivates the geometric growth of $m$: since $|\sin(4m\theta)| \le 1$, the subtracted term is bounded in magnitude by $1/(4m\sin 2\theta)$, so that $P_{\text{succ}}(m) \ge 1/4$ whenever $m \ge 1/\sin(2\theta)$. The scale $1/\sin(2\theta)\approx\tfrac12\sqrt{N/M}$ in the sparse-density regime marks the onset of this constant-success bound, not a stopping condition. Since $M$ is unknown, failed trials continue to expand $m$ up to the density-independent cap $\sqrt N$.

    \subsubsection{The Fixed-Point Search Method}
        To resolve the over-rotation inherent in standard Grover search, an alternative family of approaches, known as fixed-point quantum search algorithms, alters the fundamental construction of the phase-shift operations. Early schemes of this type, such as Grover's $\pi/3$ sequence, achieve monotonic convergence toward the target subspace but forfeit the quadratic speedup in doing so~\cite{grover2005}; the formulation adopted here instead recovers the optimal query scaling~\cite{yoder2014}.

        The structure of the fixed-point search is parameterized by an odd integer length $L = 2l + 1$, where $l$ is the number of generalized Grover iterations, a maximum allowable error tolerance $\delta \in (0, 1)$, and a minimum target density threshold $\Lambda \in (0, 1]$.
        \begin{equation}
\begin{aligned}
            L&\geq\frac{\operatorname{arccosh}(1/\delta)}
            {\operatorname{arccosh}(1/\sqrt{1-\Lambda})},\\
            \phi_j&=2\arctan\!\left[
            \frac{1}{\tanh(\operatorname{arccosh}(1/\delta)/L)
            \tan(2\pi j/L)}\right],\\
            \alpha_j&=\phi_j,\quad\beta_j=\phi_{l-j+1},\quad j=1,\ldots,l.
        \end{aligned}
\label{eq:fp_phases}
\end{equation}
        This assignment uses the positive phase exponents in both $O(\beta)$ and $D(\alpha)$ defined above. Yoder \textit{et al.}~\cite{yoder2014} use $e^{-i\alpha}$ for the initial-state reflection and $e^{i\beta}$ for the target oracle, with $\alpha_j=-\beta_{l-j+1}$. The 2 sequences are related by complex conjugation and an overall phase, yielding identical measurement probabilities for the real uniform initial state used here. Note further that the sequence is nonadaptive: once $L$, $\delta$, and $\Lambda$ are fixed, the entire phase schedule is determined in advance and requires no intermediate measurement.

        The success probability of this scheme can be modeled by exploiting the properties of Chebyshev polynomials of the first kind, denoted $T_L(x)$.
        \begin{equation}
\begin{aligned}
            \gamma&=\cosh\!\left[\operatorname{arccosh}(1/\delta)/L\right],\\
            T_L(x)&=\begin{cases}
                \cos(L\arccos x),&|x|\leq1,\\
                \cosh(L\operatorname{arccosh}x),&x>1,
            \end{cases}\\
            P_{\mathrm{FP}}(l,\delta,a)&=1-\delta^2T_L^2(\gamma\cos\theta),\\
            a\geq\Lambda&\ \Longrightarrow\ \gamma\cos\theta\leq1
            \ \Longrightarrow\ P_{\mathrm{FP}}(l,\delta,a)\geq1-\delta^2.
        \end{aligned}
\label{eq:fp_success_probability}
\end{equation}
        If $\gamma\cos\theta > 1$, which can occur when the true density falls below the assumed threshold, the hyperbolic branch applies and the guarantee need not hold. The scheme is therefore robust to any underestimate of the solution density but not to an overestimate, which is the essential trade-off distinguishing it from the adaptive BBHT approach.

    \subsubsection{Density Estimation Methods}

        DEM includes classical and quantum techniques for estimating $a$, from which the target count is recovered as $M=aN$. Examples include Quantum Counting~\cite{brassard1998,brassard2002}, which combines Grover iterations with Fourier-based phase estimation; Iterative Quantum Amplitude Estimation~\cite{grinko2021}, which uses variable-depth circuits and classical updates to narrow density confidence intervals; and Maximum Likelihood Amplitude Estimation~\cite{suzuki2020}, which fits the density to measurement outcomes. Classical WL sampling~\cite{wang2001} instead estimates the density of objective values and the fraction satisfying a search threshold. This option is used in the potential application below.

\section{Optimization Formulation}

The analyses presented above characterize the success probability
of each method as a function of its parameters, but do not directly
answer the practical cost question: how many oracle queries are
expended, in expectation, before a solution is first observed?
The asymptotic results accompanying the original proposals~
\cite{boyer1998,yoder2014} do not by themselves determine which
method has the lowest expected cost at a particular density.
This section formulates that comparison as an optimization over
the user-controlled parameters of each method.

\subsection{Query Metric and Parameter Sets}

Let $\omega$ denote the user-controlled parameters. BBHT optimizes its expansion
factor $\lambda$; FP selects its sequence depth $l$ and error
parameter $\delta$ under a prescribed density threshold $\Lambda$;
and DEM abstracts the estimation stage, leaving the standard Grover
iteration count $k$ to be optimized. Each optimization is performed
independently at each benchmark density under both initialization-cost
conventions: $c=0$ excludes state-preparation cost, whereas $c=1$
charges 1 additional unit per preparation, as detailed in the next subsection.

Retaining the preceding notation with $a>0$, for a fixed trial configuration of method
$F\in\{\mathrm{BBHT},\mathrm{FP},\mathrm{DEM}\}$, let $x_F$ be
its query count and $P_F(x_F,\omega,a)$ its success probability.
Independent repetition of the same configuration until first
success gives the expected cost
\begin{equation}
 \mathrm{query}_{F,a,c}(\omega)
 =\frac{x_F+c}{P_F(x_F,\omega,a)},
 \label{eq3.1}
\end{equation}
where $c\in\{0,1\}$ assigns either 0 or 1 additional cost unit
to each preparation of the uniform state. Configurations with success probability 0 have infinite expected cost.

For FP, $x_F=l$; for idealized DEM, $x_F=k$. BBHT randomizes its
iteration count and changes its sampling bound after failure, so
its total expected cost requires the survival-weighted extension
in \eqref{eq3.1.1}. The objective in each case is to find the
configuration $\omega^*_{F,a,c}$ minimizing the corresponding cost.
Reported optima refer to
the candidates tested, subject to numerical evaluation error;
grid spacing alone does not certify closeness to an unrestricted
continuous-domain optimum.

The benchmarks use $N=2^{50}$ and 3 overlapping target-density
grids: $[10^{-3},0.499]$ with spacing $10^{-3}$,
$[10^{-4},10^{-2}]$ with spacing $10^{-4}$, and
$[10^{-5},10^{-3}]$ with spacing $10^{-5}$.
At each target value, $M=\lfloor a_{\mathrm{target}}N\rfloor$ and
the resulting integer target count is used in all cost evaluations.

\subsection{Initialization-Cost Conventions}

The 2 initialization-cost conventions, $c=0$ and $c=1$, distinguish minimizing oracle usage from accounting for repeated state preparation. The first, $c=0$, isolates
search-oracle usage~\cite{nielsen2010,brassard2002}. The second,
$c=1$, includes 1 additional unit per preparation and permits
the zero-iteration baseline, whose expected cost is $1/a$.
This provides a complementary measure of repeated-execution cost;
the assigned unit is a convention rather than
an equality between the physical gate costs of preparation and
an oracle. One application of $O(\beta)$ is counted as 1 search query for every method. Quantum costs are reported in query-cost units; gate counts and circuit execution times are not analyzed. Yoder \textit{et al.}~\cite{yoder2014} count $2l$ underlying Boolean-marker calls per FP sequence; the conversion depends on implementation~\cite{nagy2024}.

When $c=0$, repeating 0 Grover iterations yields 0 charged
search cost at every positive density. Such sampling still requires
state preparation, measurement, and verification, and would make
the search-only optimization degenerate. Therefore, $l\geq1$ for FP and $k\geq1$ for DEM are required
under this convention.
Zero-iteration draws remain part of BBHT's expanding randomized
schedule, which is optimized as a whole.

The numerical comparison focuses on $0<a<1/2$. For $a\geq1/2$,
uniform random sampling costs $1/a\leq2$ under $c=1$ and is no
more costly than a fixed configuration using at least 1 query.
For an integer standard-Grover count $x\geq1$, the stronger
oracle-count comparison follows directly from
$\theta=\arcsin\sqrt a$:
\begin{equation}
\begin{aligned}
 x=1:\quad&\sin^2(3\theta)=a(3-4a)^2\leq a,\\
 x\geq2:\quad&\sin^2((2x+1)\theta)\leq1\leq ax,\\
 &\frac{x+1}{\sin^2((2x+1)\theta)}
 \geq\frac{x}{\sin^2((2x+1)\theta)}\geq\frac1a.
\end{aligned}
\label{eq:high_density_comparison}
\end{equation}
Here $(3-4a)^2\leq1$ for $1/2\leq a\leq1$.
Under $c=0$, zero-query sampling is excluded for the reason above.

\subsection{Optimization for Boyer--Brassard--H{\o}yer--Tapp Search}

The expansion factor $\omega=\lambda>1$ controls the growth of the sampling
bound. The benchmark uses a variant of the previously described
schedule: its cap is based on the first Grover peak for 1 marked
state, replacing the $\sqrt N$ cap in the introductory algorithm:
\begin{equation}
\begin{aligned}
 m_1&=1,\qquad m_{s+1}=\min\{\lambda m_s,m_{\max}\},\\
 m_{\max}&=\max\!\left\{1,
 \left\lfloor\frac{\pi}{4\arcsin\sqrt{1/N}}\right\rfloor\right\}.
\end{aligned}
\label{eq:bbht_capped_schedule}
\end{equation}
This is a schedule cap, not a fundamental limit on executable
Grover iterations. At trial $s$, let $J_s=\lceil m_s\rceil$.
Uniformly sampling $k_s\in\{0,\ldots,J_s-1\}$ gives
$\mathbb E[k_s]=(J_s-1)/2$. Write
$P_s=P_{\mathrm{succ}}(J_s)$ using the average success probability
established in the BBHT section, and let
$p_s=\prod_{i=1}^{s-1}(1-P_i)$ be the probability of reaching
trial $s$, with $p_1=1$. The total expected cost is then
\begin{equation}
 \mathrm{query}_{\mathrm{BBHT},a,c}(N,\lambda)
 =\sum_{s=1}^{\infty}p_s
       \left(\frac{J_s-1}{2}+c\right).
 \label{eq3.1.1}
\end{equation}
The survival weights account for all unsuccessful trials and for
the changing cost and success probability along the schedule.
The parameter optimization is therefore
\begin{equation}
 \lambda^*_{a,c}\in
 \underset{\lambda\in\mathcal L}{\arg\min}
 \mathrm{query}_{\mathrm{BBHT},a,c}(N,\lambda),
 \label{eq:bbht_parameter_optimization}
\end{equation}
where $\mathcal L=\{1+10^{-5},1+2\times10^{-5},\ldots,6\}$
is the numerical sweep grid. Thus $\omega^*_{\mathrm{BBHT},a,c}
=\lambda^*_{a,c}$. The ceiling in $J_s$ makes the objective
nonsmooth, motivating evaluation by a discrete sweep.

To evaluate \eqref{eq3.1.1}, let $s_*$ be the first trial reaching
$m_{\max}$, and define $d_{c,s}=(J_s-1)/2+c$. Subsequent trials
repeat the same configuration, giving the exact geometric tail
\begin{equation}
 \mathrm{query}_{\mathrm{BBHT},a,c}(N,\lambda)
 =\sum_{s=1}^{s_*-1}p_s d_{c,s}
  +p_{s_*}\frac{d_{c,s_*}}{P_{s_*}}.
 \label{eq:bbht_exact_tail}
\end{equation}
Since the capped distribution includes $k=0$,
$P_{s_*}\geq a/\lceil m_{\max}\rceil>0$.
Consequently, this expectation is finite for fixed $N$, $a>0$,
and every $\lambda>1$.

\subsection{Optimization for Fixed-Point Search}

$x_{\mathrm{FP}}=l$ and the optimized parameters are
$\omega=(l,\delta)$. The threshold $\Lambda\leq a$ is prescribed
by the information scenario. Substituting the previously
established FP success probability into \eqref{eq3.1} gives
\begin{equation}
 \mathrm{query}_{\mathrm{FP},a,c}(l,\delta)
 =\frac{l+c}{P_{\mathrm{FP}}(l,\delta,a)}.
 \label{eq:fp_expected_query}
\end{equation}

Related expected-query optimization for FP search was studied
by Nagy \textit{et al.}~\cite{nagy2024}, who numerically optimized
the known-density cost ratio in the sparse-density limit and
developed an increasing-depth strategy for unknown density.
Here, the density-specific configurations over the
benchmark range are evaluated under both initialization conventions,
$c=0$ and $c=1$, using the phase-selection and depth-search
rules below.

Restricting $\delta$ to the tested grid of $10^6$ equally spaced
points between $10^{-10}$ and $1-10^{-15}$, the FP length condition
gives the admissible set
\begin{equation}
 \mathcal D_l(\Lambda)
 =\left\{\delta:
 \delta\geq\frac{1}{
 \cosh\!\left[(2l+1)\operatorname{arccosh}
 \!\left(\frac{1}{\sqrt{1-\Lambda}}\right)\right]}
 \right\}.
 \label{eq:fp_admissible_delta_grid}
\end{equation}
The 2 scenarios are considered.
\begin{equation}
\begin{aligned}
 \Lambda=1/N:\quad&
 \delta_l=\min\mathcal D_l(1/N),\\
 \Lambda=a:\quad&
 \delta_l\in\underset{\delta\in\mathcal D_l(a)}{\arg\max}
 P_{\mathrm{FP}}(l,\delta,a).
\end{aligned}
\label{eq:fp_delta_selection}
\end{equation}
At fixed $l$, maximizing the success probability minimizes expected cost
because $l+c$ is constant. The rule for $\Lambda=1/N$ instead optimizes
the threshold guarantee, which need not maximize success at the
actual density.
For the reported benchmark settings, the grid contains an
admissible value at every positive depth. If the admissible
grid is empty for other settings, the implementation raises
an error, requiring the grid to be extended before proceeding.
In both scenarios, the outer depth optimization uses the true
benchmark density. Consequently, $\Lambda=1/N$ prescribes the
phase-selection rule, not a density-independent procedure
for selecting the best depth.

When $c=1$, the search includes the zero-depth baseline,
whose success probability is $a$ and expected cost is $1/a$.
For this phase-free case, set $\delta_0=0$ and evaluate its cost
directly as $1/a$. When $c=0$, only positive depths are allowed.
The depth search is given in Algorithm~\ref{alg:configuration_search}.
Since $P_{\mathrm{FP}}\leq1$, every configuration at depth $l$
costs at least $l+c$; the stopping condition therefore excludes
all larger depths from improving the stored result.
The returned depth and phase satisfy
\begin{equation}
 \begin{aligned}
  l^*_{a,c}&\in\underset{l\in\mathbb Z_{\geq1-c}}{\arg\min}
        \mathrm{query}_{\mathrm{FP},a,c}(l,\delta_l),\\
  \delta^*_{a,c}&=\delta_{l^*_{a,c}}.
 \end{aligned}
 \label{eq:fp_outer_optimization}
\end{equation}
Phase resolution remains limited by the stated grid and
selection rules.

\begin{algorithm}[!t]
\caption{FP and DEM configuration search}
\label{alg:configuration_search}
\begin{algorithmic}[1]
\Require Method $F\in\{\mathrm{FP},\mathrm{DEM}\}$, $N$, $a$, $c$;
FP scenario and grid when $F=\mathrm{FP}$
\Ensure Stored configuration $\omega^*$ and best expected cost $B$
\If{$F=\mathrm{FP}$}
    \State $l\gets1$, $B\gets+\infty$
    \State If $c=1$, set $B\gets1/a$, $\omega^*\gets(0,0)$
\Else
    \State $k\gets1-c$, $k_{\mathrm p}\gets\max\{1-c,\lfloor\pi/(4\theta)\rfloor\}$
    \State $B\gets\mathrm{query}_{\mathrm{DEM},a,c}(k_{\mathrm p})$, $\omega^*\gets k_{\mathrm p}$
\EndIf
\While{$x_F+c\leq B$, with $x_F=l$ for FP or $k$ for DEM}
    \State For FP, select $\delta_l$ by \eqref{eq:fp_delta_selection}; error if the grid is empty
    \State $\omega\gets(l,\delta_l)$ for FP, or $k$ for DEM
    \State $q\gets\mathrm{query}_{F,a,c}(\omega)$
    \If{$q<\infty$ and ($q<B$ or an equal-cost configuration uses fewer iterations)}
        \State $B\gets q$, $\omega^*\gets\omega$
    \EndIf
    \State Increment $l$ for FP, or $k$ for DEM
\EndWhile
\State \Return stored $\omega^*$ and $B$
\end{algorithmic}
\end{algorithm}

\subsection{Optimization for Density Estimation Methods}

The estimation methods described earlier incur costs that depend
on the estimator, accuracy, and confidence requirements. To keep
the benchmark independent of that choice, estimation is abstracted
and standard Grover search is evaluated with exactly known $a$. The only optimized parameter is $\omega=k$, with $x_{\mathrm{DEM}}=k$.
Using the standard Grover success probability, the objective becomes
\begin{equation}
\begin{aligned}
 \mathrm{query}_{\mathrm{DEM},a,c}(k)
 &=\frac{k+c}{\sin^2((2k+1)\theta)},\\
 k_c^*(a)&\in\underset{k\in\mathbb Z_{\geq1-c}}{\arg\min}
 \mathrm{query}_{\mathrm{DEM},a,c}(k).
\end{aligned}
\label{eq3.2}
\end{equation}
The search is given in Algorithm~\ref{alg:configuration_search}.
Since success probability is at most 1, every configuration
costs at least $k+c$. All larger counts also have cost exceeding
$B$, so this stopping rule excludes no improving configuration.
With $k=0$ admitted only for $c=1$, the returned count gives
$\omega^*_{\mathrm{DEM},a,c}=k_c^*(a)$ and minimizes expected
cost to first success over the admissible integer counts.

An actual estimator
adds computation and can change the selected $k$ through estimation
error, so its deviation from this reference need not equal the
estimation overhead alone. In the adaptive framework, DEM reuses
the outer WL density estimate when selected.

The benchmark tables record the optimized configuration and expected
cost of each method under each convention. Both FP threshold cases, $\Lambda=1/N$ and $\Lambda=a$, are included as candidate configurations. Comparing these costs
identifies the winning method at each density; its parameters and
cost may vary even within an interval where the winning method
remains unchanged. Table construction is reusable offline
preprocessing, distinct from the online execution cost.

\section{Numerical Benchmark Results}

\begin{figure*}[!t]
    \centering
    \includegraphics[width=\linewidth]{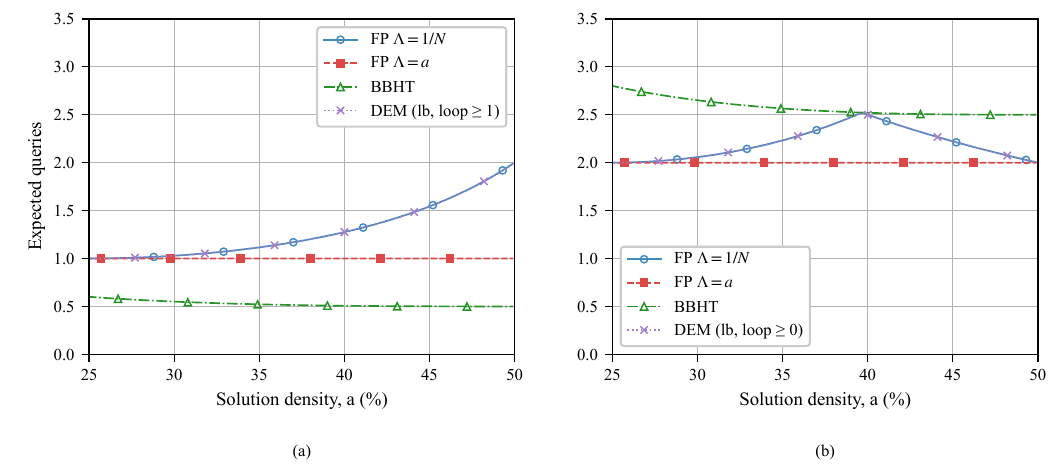}
    \caption{Optimal queries on selected candidate methods for each solution density $a\in[0.25,0.499]$ where (a) shows the case when the Hadamard layer is assigned 0 cost ($c=0$), whereas (b) shows when counted as 1 query ($c=1$).}
    \label{fig5.1}
\end{figure*}

\begin{figure*}[!t]
    \centering
    \includegraphics[width=\linewidth]{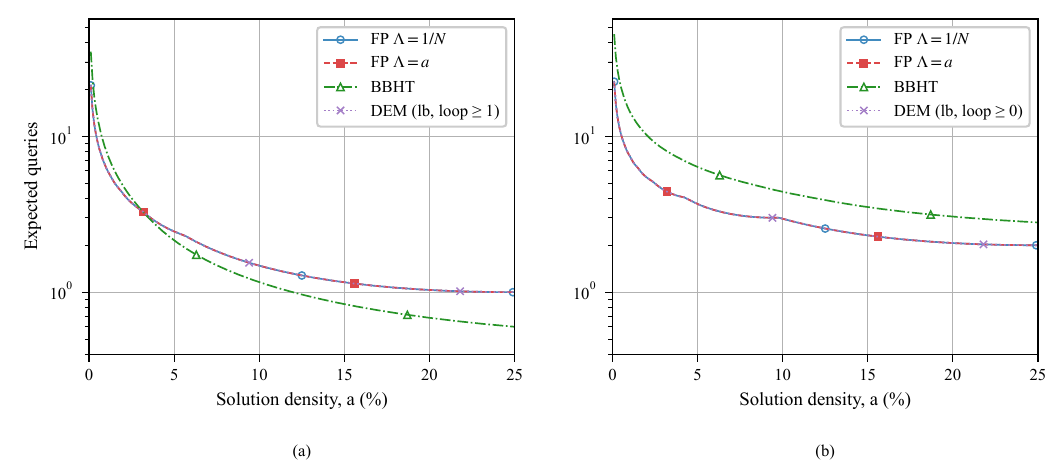}
    \caption{Optimal queries (in log scale) on selected candidate methods for each solution density $a\in[0.001,0.25]$ where (a) shows the case when the Hadamard layer is assigned 0 cost ($c=0$), whereas (b) shows when counted as 1 query ($c=1$).}
    \label{fig5.2}
\end{figure*}

\begin{figure*}[!t]
    \centering
    \includegraphics[width=\linewidth]{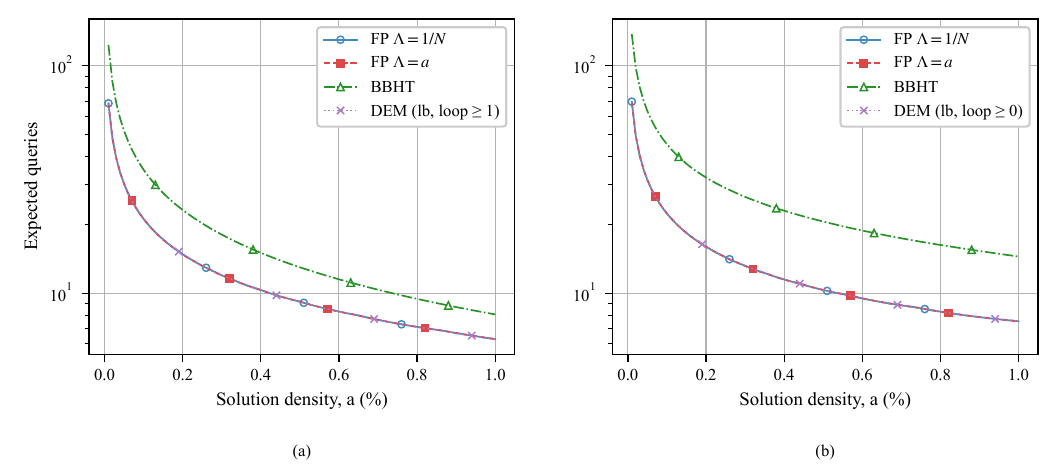}
    \caption{Optimal queries (in log scale) on selected candidate methods for each solution density $a\in[10^{-4},0.01]$ where (a) shows the case when the Hadamard layer is assigned 0 cost ($c=0$), whereas (b) shows when counted as 1 query ($c=1$).}
    \label{fig5.3}
\end{figure*}

\begin{figure*}[!t]
    \centering
    \includegraphics[width=\linewidth]{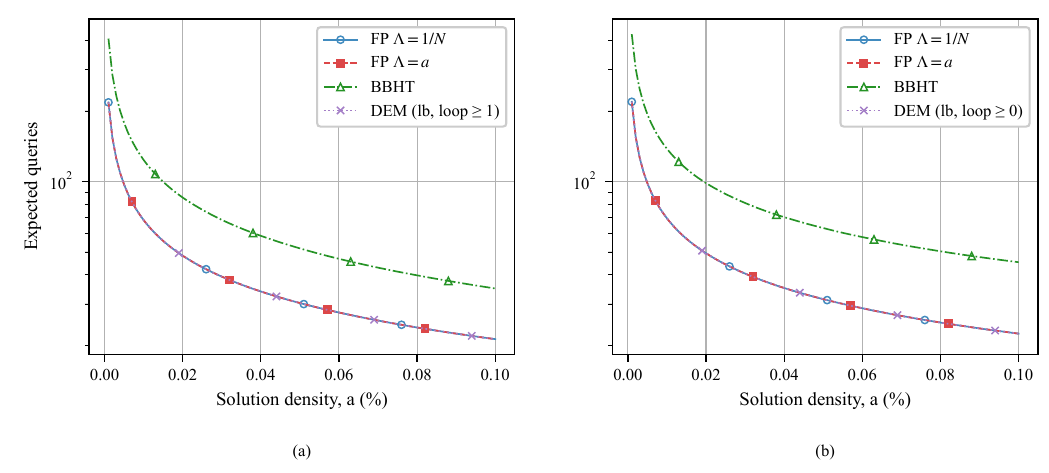}
    \caption{Optimal queries (in log scale) on selected candidate methods for each solution density $a\in[10^{-5},10^{-3}]$ where (a) shows the case when the Hadamard layer is assigned 0 cost ($c=0$), whereas (b) shows when counted as 1 query ($c=1$).}
    \label{fig5.4}
\end{figure*}

Figs.~\ref{fig5.1}--\ref{fig5.4} compare the optimized costs on the tested density grids. At $a\in[0.25,0.499]$, BBHT requires approximately 0.5 queries when $c=0$, whereas FP with $\Lambda=a$ gives the lowest cost, approximately 2 queries, when $c=1$. For $c=0$, BBHT remains preferable at 3.1\% density but is outperformed at 3.0\%; below this sampled transition, FP and DEM have numerically indistinguishable costs and outperform BBHT. For $c=1$, FP and DEM are already preferable below 25\% density. Across the finer grids down to $a=10^{-5}$, their query gap from BBHT grows as density decreases.

The high-density behavior also depends on the FP threshold. For $\Lambda=1/N$, the $c=0$ cost decreases from approximately 2 queries at 49.9\% density to 1 query. Under $c=1$, it instead rises to approximately 2.5 queries near 39.7\%, where the zero-depth baseline gives way to positive depths, before falling to approximately 2 queries at 27.9\%. DEM follows the same numerical cost curve. The gap from FP with $\Lambda=a$ falls below 1\% near 27.9\%, indicating that the threshold choice primarily affects the denser regimes tested. Minor fluctuations near 9.8\% accompany transitions from $l^*=1$ to larger depths. Below 0.004\% density, initialization accounting changes the FP and DEM costs by less than 1\%. 


\section{Application to Grover Adaptive Search}

    Quantum approximate optimization~\cite{farhi2014}, variational quantum eigensolvers~\cite{peruzzo2014}, and Quantum Annealing~\cite{kadowaki1998} are heuristic optimization techniques. GAS provides a framework for exact optimization~\mbox{\cite{durr1996,gilliam2021}} and is used below as a potential application of the benchmarks.

        Specifically, GAS dynamically redefines the marked states to search for improving solutions. For minimization of $f(x)$ over an $n$-qubit search space, Algorithm~\ref{alg:gas} repeats until the target set is empty ($M_y=0$); an exact empty-target test certifies the final incumbent as globally optimal.

        \begin{algorithm}[t]
            \caption{GAS}
            \label{alg:gas}
            \begin{algorithmic}[1]
                \Require Objective $f: \{0,1\}^n \to \mathbb{R}$, iteration-selection rule
                                \State Select random initial state $x_0 \in \{0,1\}^n$
                \State Set initial threshold $y \gets f(x_0)$
                \State Set incumbent $x_{\mathrm{best}} \gets x_0$
                \Repeat
                    \State $M_y \gets |\{x \mid f(x) < y\}|$
                    \If{$M_y=0$}
                        \State \textbf{break}
                    \EndIf
                    \State $g_y(x) \gets \begin{cases} 1 & \text{if } f(x) < y \\ 0 & \text{otherwise} \end{cases}$
                    \State Construct Grover Operator $G_y$ satisfying $g_y(x)$
                    \State Select $k$ using the supplied rule\label{alg:gas_iteration}
                    \State $|\psi_0\rangle \gets H^{\otimes n}|0\rangle^{\otimes n}$
                    \State $|\psi_k\rangle \gets (G_y)^k |\psi_0\rangle$
                    \State $x^* \gets \text{Measure}(|\psi_k\rangle)$
                    \If{$f(x^*) < y$}
                        \State $x_{\mathrm{best}} \gets x^*$
                        \State $y \gets f(x_{\mathrm{best}})$
                    \EndIf
                \Until{$M_y = 0$}
                \State \Return Global optimum state $x_{\mathrm{best}}$ and minimum value $y$
            \end{algorithmic}
        \end{algorithm}

        To use the numerical benchmarks within GAS, the target count $M_y$ at the current threshold must be estimated. BBHT handles an unknown count indirectly through randomized iteration counts, whereas FP uses a prescribed density threshold to construct its phase sequence. DEM estimates the count directly using a classical or quantum subroutine. In the potential application considered here, WL is used to estimate $M_y$ and select a method and configuration from the benchmark tables.

        To estimate the density of states $g(E)$ across an unmapped $n$-qubit discrete optimization landscape $\{0,1\}^n$ without exhaustive enumeration, the algorithm executes a flat-histogram random walk in configuration space. Each invocation has a user-specified budget of $S\geq1$ trial steps.
        The core WL transition mechanics are retained, whereas unknown objective bounds are handled through dynamic score-bin discovery. Algorithm~\ref{alg:wl_density} initializes the first bin and allocates an entry whenever a new proposed score is encountered. Consequently, the flatness criterion in Algorithm~\ref{alg:wl_density} evaluates strictly over the \textit{discovered} score bins rather than the complete score range; $\bar H$ denotes their mean histogram count.
        The estimated density and target count are normalized via
        \begin{equation}
        \begin{aligned}
            g(E)&=N\frac{e^{\ln g(E)-\ln g_{\max}}}
                        {\sum_{E'}e^{\ln g(E')-\ln g_{\max}}},\\
            \hat a&=\frac{\sum_{E\geq y}g(E)}{N},\qquad
            M_y\approx\hat M_y=N\hat a.
        \end{aligned}
        \label{eq:wl_density_output}
        \end{equation}

        \begin{algorithm}[!t]
        \caption{WL density estimation with dynamic score bins}
        \label{alg:wl_density}
        \begin{algorithmic}[1]
        \Require $n$, objective $f$, threshold $y$, budget $S$, check interval $b$, flatness factor $\gamma$, initial $\ln f>0$, tolerance $\epsilon$
        \Ensure $\hat M_y$, $\hat a$
        \State Sample $x\in\{0,1\}^n$ uniformly; $E\gets f(x)$
        \State Initialize $\ln g(E)\gets0$, $H(E)\gets0$
        \For{$s=1,\ldots,S$}
            \State Flip 1 uniformly chosen bit of $x$ to propose $x'$; $E'\gets f(x')$
            \State If $E'$ is new, initialize $\ln g(E')\gets0$, $H(E')\gets0$
            \State With probability $\min(1,e^{\ln g(E)-\ln g(E')})$, set $(x,E)\gets(x',E')$
            \State $H(E)\gets H(E)+1$; $\ln g(E)\gets\ln g(E)+\ln f$
            \If{$s$ is a multiple of $b$ and $\min_E H(E)>\gamma\bar H$}
                \State Reset all $H(E)\gets0$; $\ln f\gets\frac12\ln f$
                \If{$\ln f<\epsilon$} \State \textbf{break} \EndIf
            \EndIf
        \EndFor
        \State $\ln g_{\max}\gets\max_E\ln g(E)$
        \State Normalize and calculate $\hat M_y$, $\hat a$ using \eqref{eq:wl_density_output}
        \State \Return $\hat M_y$, $\hat a$
        \end{algorithmic}
        \end{algorithm}

\begin{figure*}[!t]
    \centering
    \includegraphics[width=\linewidth]{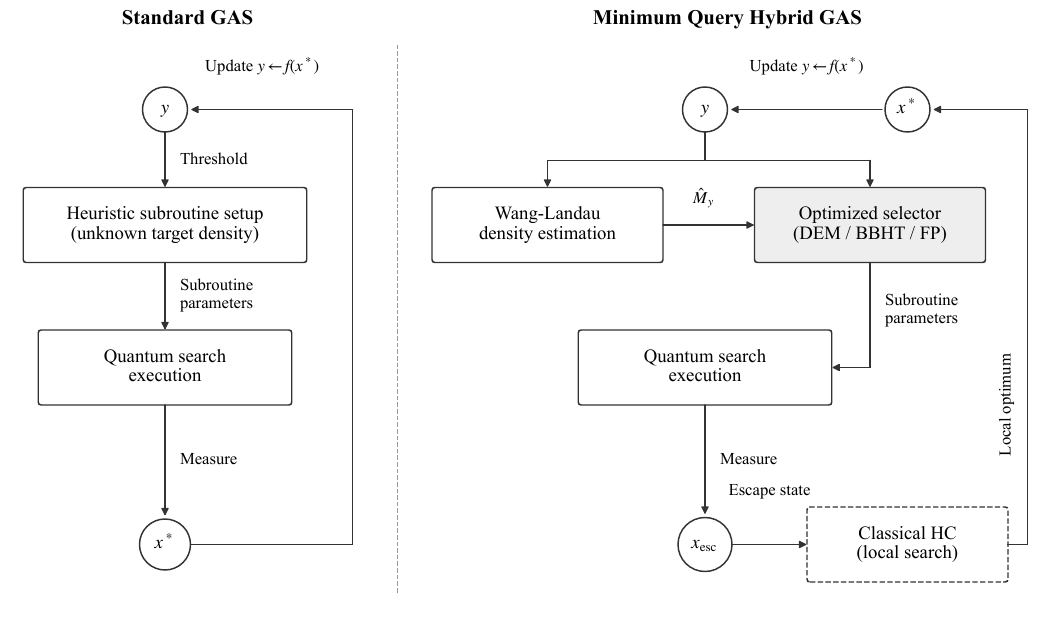}
    \caption{Comparison of the Standard GAS architecture versus the proposed Hybrid GAS architecture. Standard GAS (left) operates as a generalized framework, applying an arbitrary quantum search subroutine to iteratively find improving states. The proposed method (right) enhances this framework by utilizing a Wang-Landau density estimation subroutine to approximate the target count $\hat{M}_y$. This estimate feeds into an optimized selector that dynamically routes execution to the selected quantum approach (BBHT, FP, DEM). The measured quantum state $x_{\text{esc}}$ acts strictly as an inter-basin escape mechanism, handing execution off to a classical HC algorithm for intra-basin exploitation.}
    \label{fig6.1}
\end{figure*}

        Hybrid GAS for continuous optimization was provided in a prior study~\cite{lara2014new}. In the resulting use-case, a classical local search is inserted between quantum calls; HC is used here to establish a local improvement threshold. In this potential use-case, WL estimates the corresponding target fraction, the benchmark selector supplies the quantum method and configuration, and a successful sample starts the next HC invocation. Fig.~\ref{fig6.1} compares this example with traditional GAS.

\section{Experimental Methodology}
\subsection{Cumulative Expected Query Analysis}
    To evaluate the overall efficiency of the hybrid quantum-classical search, the algorithm's progression through the problem landscape is modeled as an absorbing Markov chain. Let $o_1<o_2<\cdots<o_K$ denote the distinct local-optimum fitness values obtained by the hill-climber. Let $B_k$ contain all starting states whose hill-climbing endpoint has fitness $o_k$, and let $S_i$ contain all states with fitness strictly greater than $o_i$.
    \begin{equation}
\begin{aligned}
        \pi_k&=\frac{|B_k|}{N},\qquad M_i=|S_i|,\\
        P_{i,j}&=\frac{|S_i\cap B_j|}{M_i},\qquad i<K.
    \end{aligned}
\label{eq:cumulative_transition_counts}
\end{equation}
    These parameters are extracted directly via exact computational enumeration of the finite state space. For each fixed cost convention $c$, the method $F$ and configuration $\omega$ at a basin are prescribed by the test configuration or selected from the benchmark tables. The single-stage expected cost $Q_i$ uses \eqref{eq3.1.1} for BBHT, \eqref{eq:fp_expected_query} for FP, or \eqref{eq3.2} for DEM, evaluated at the true density $M_i/N$. The cumulative expected cost $E_i$ is the immediate quantum query cost to escape the current basin plus the expected remaining cost evaluated over all possible future basins $j$:
    \begin{equation}\label{eq:8.2}
    \begin{aligned}
        Q_i&=\mathrm{query}_{F,M_i/N,c}(\omega),\qquad i<K,\\
        E_K&=0,\\
        E_i&=Q_i+\sum_{j=i+1}^{K}P_{i,j}E_j,\qquad i=K-1,\ldots,1.
    \end{aligned}
    \end{equation}
    Here, $P_{i,j}$ is the transition probability of the hill-climber landing in basin $j$ given that a successful quantum sample was found. Because the hybrid algorithm strictly progresses to higher fitness values, $P_{i,j}=0$ for $j\leq i$. The total expected quantum query cost $E_{{global}}$, for reaching the global optimum from a random initialization, is then the weighted sum of the expected costs to finish the search from any initial starting basin $k$:
    \begin{equation}\label{eq:8.1}
        E_{{global}}=\sum_{k=1}^{K}\pi_k E_k.
    \end{equation}

\subsection{Experimental Configuration}
    The framework is tested on combinatorial optimization problems. The metric is the accumulated quantum query count until the global optimum is found. The starting points are sampled randomly. The trap framework stops when $\hat{M}_y=0$. For the knapsack framework and the BBHT/FP baselines, the stopping test uses the exactly enumerated condition $M_y=0$. 

    Nevertheless, the search beyond the optimal point, in which $M_y=0$, requires theoretically infinite queries, because a better solution does not exist. All states in both test instances are enumerated to verify the reported terminal solutions. A more robust termination condition is recommended for practical implementations and further studies in different setups and scales, as the Wang-Landau algorithm might have made a mistake in identifying the global optimum. 

    The demonstrations are conducted in 3 cases; all of them are executed using the hybrid GAS, based on hill climbing. The first case uses the BBHT algorithm at $\lambda=1.2$, the second case uses FP at the minimum $l$ satisfying \eqref{eq:fp_phases} for $\delta=0.999$ and $\Lambda=1/N$, and the third case is the framework that uses the Wang-Landau algorithm with the method and configuration selected from the benchmarks. It is important to note that the setups for the BBHT and FP cases are 2 possible nonoptimized configurations, used purely as naive examples, meaning that there might exist better parameters that yield smaller accumulated queries to the global optimum. DEM uses $\hat a=0.7a$ to select $k$ from the nearest DEM benchmark density; \eqref{eq3.2} evaluates its cost at the true density $a$. No particular estimator is instantiated in this reference. Aer implements WL within the adaptive framework, including its DEM branch. The analytical framework uses $\hat a=a$. For WL, a maximum of $S=\lfloor N/2\rfloor$ steps per invocation is used, with initial $\ln f=0.03$, $\gamma=0.6$, $b=64$, and $\epsilon=10^{-6}$. For the framework, the benchmark entry nearest to the WL estimate is selected. Such selections aim to provide near-optimal performance at the queried density. Each case covers 2 perspectives, including when $c=0$ (H layer initialization is cost-free) and when $c=1$ (H layer initialization is counted as 1 query), according to \eqref{eq3.1}.

\subsection{Test Landscapes}
    There are 2 binary optimization instances used in this demonstration. First, the deceptive trap serves as a toy problem that is hard for traditional hill-climbing. Second, the 0/1 Knapsack problem is used as a constrained test instance.

    \subsubsection{Deceptive Trap}
        The deceptive trap instance is constructed as a custom landscape inspired by Deb and Goldberg\textquotesingle s analysis of trap functions~\cite{deb1993analyzing}, designed to mislead heuristic and local search algorithms. The landscape evaluates an input bitstring sequentially by dividing it into nonoverlapping 4-bit blocks, combining a linear slope reward with a strict conditional trap modifier.

        \begin{equation}
\begin{aligned}
            \max \quad & \sum_{i=0}^{k^*} \left[10 (b_{i,0} + b_{i,1}) + \Delta_i\right] \\
            \text{where} \quad & \Delta_i = \begin{cases}
            50 & \text{if } b_{i,2} + b_{i,3} = 2 \\ 
            -50 & \text{if } b_{i,2} + b_{i,3} = 1 \\ 
            0 & \text{if } b_{i,2} + b_{i,3} = 0 
            \end{cases} \\
            \text{s.t.} \quad & b_{i,j} \in \{0, 1\}.
        \end{aligned}
\label{eq:trap_objective}
\end{equation}

        $b_{i,j} \in \{0, 1\}$ represents the $j$-th binary decision variable within block $i$. Within each block, the score consists of 2 components: a baseline linear reward $10(b_{i,0} + b_{i,1})$ derived from the first 2 "slope" bits, and a conditional modifier $\Delta_i$ determined by the remaining 2 "trap" bits. The parameter $k^*$ represents the index of the first block where the optimal trap condition is not satisfied ($b_{i,2} + b_{i,3} < 2$), or the last block when every block satisfies that condition. This acts as a truncation point for the objective function. Any variables located in blocks beyond $k^*$ do not contribute to the total score, effectively flattening their local gradients to 0. 

\begin{figure*}[!t]
    \centering
    \includegraphics[width=\linewidth]{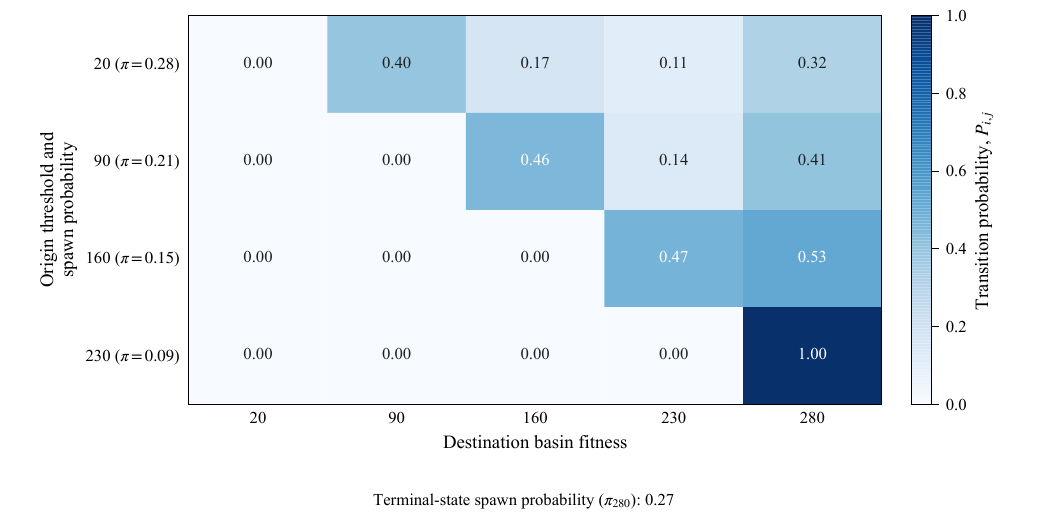}
    \caption{Landscape information of the test deceptive trap instance. The heatmap represents the hill-climbing heuristic's transition probabilities from origin basins (y-axis) to a destination basin (x-axis) across all possible solution states. Here, $\pi$ on the y-axis represents the chance of spawning in a certain basin (the chance to spawn at the global optimum is noted at the bottom).}
    \label{fig9.1}
\end{figure*}

        This formulation introduces a dual-layered deceptive landscape. First, transitioning from a neutral trap state ($b_{i,2} + b_{i,3} = 0$) to the globally optimal state ($b_{i,2} + b_{i,3} = 2$) requires traversing an intermediate state that triggers a severe penalty of $\Delta_i = -50$ in \eqref{eq:trap_objective}. Second, the truncation at $k^*$ in \eqref{eq:trap_objective} masks the objective contribution of downstream variables, depriving heuristic routines of the gradient information necessary to escape premature local minima. In this study, a landscape of 4 block sets is implemented, equal to 16 binary decision variables, or $2^{16}$ possible solutions. With $\pi_k$, $P_{i,j}$, and $M_i$ exhaustively calculated, Fig.~\ref{fig9.1} illustrates an overview of the landscape information that is required to compute the expected queries.

    \subsubsection{0/1 Knapsack}
        The 0/1 Knapsack dataset is constructed following Pisinger's benchmark generation framework~\cite{pisinger2005} to evaluate local search and heuristic algorithms on constrained binary hypercubes. The landscape evaluates an input bitstring $x \in \{0, 1\}^n$ by maximizing total accumulated profit subject to a strict capacity threshold:

        \begin{equation}
\begin{aligned}
                \max \quad & \sum_{i=1}^{n} p_i x_i \\
                \text{s.t.} \quad & \sum_{i=1}^{n} w_i x_i \le C, \quad x_i \in \{0, 1\}.
            \end{aligned}
\label{eq:knapsack_objective}
\end{equation}

\begin{figure*}[!t]
    \centering
    \includegraphics[width=\linewidth]{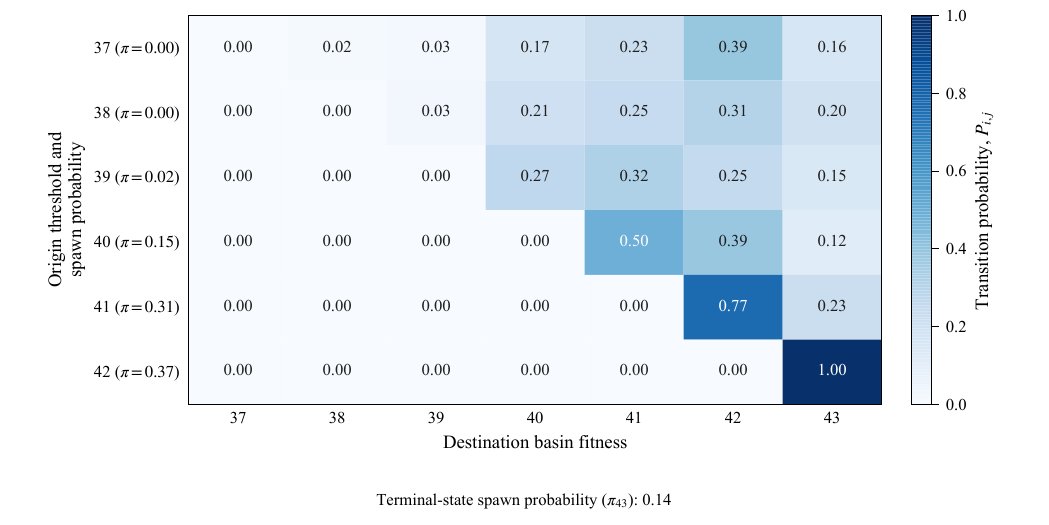}
    \caption{Landscape information of the test 0/1 Knapsack instance. The heatmap represents the hill-climbing heuristic's transition probabilities from origin basins (y-axis) to a destination basin (x-axis) across all possible solution states. Here, $\pi$ on the y-axis represents the chance of spawning in a certain basin (the chance to spawn at the global optimum is noted at the bottom).}
    \label{fig9.2}
\end{figure*}

        During HC, an infeasible bitstring is assigned fitness $\sum_i p_i x_i-[10^4+100(\sum_i w_i x_i-C)]$. During WL sampling, all infeasible bitstrings are instead assigned score $-1$; feasible bitstrings retain their profit in both procedures. The quantum oracle marks only feasible bitstrings satisfying the profit threshold. In this study, the knapsack capacity is parameterized as $C = \left\lfloor 0.5 \sum_{i=1}^{n} w_i \right\rfloor$. The problem instance is constructed using $n = 12$ decision variables, an integer weight bound $R = 10$, and a fixed seed of 2 under the strongly correlated setting, where item profits are deterministically tied to weights via $p_i = w_i + 1$ for $w_i \in \{1, \dots, R\}$. Similarly to the Deceptive Trap case, $\pi_k$, $P_{i,j}$, and $M_i$ are exhaustively calculated and are presented in Fig.~\ref{fig9.2}.

        Note that for both experiment cases, the range of the cost function is defined to be a set of integers only. Then, for both analytical methods and the real numerical executions, the aim for the search is defined as $f(x) \leq y - 1$ instead of $f(x) < y$, for the sake of simpler implementations. And since both test landscapes are for maximization, the searching goal, $f(x) \geq y + 1$, is used in this study.

\section{Numerical Experiment Results}
    Results from 300 randomly spawned points are presented, with HC + GAS executed until the global optimum is obtained. Cumulative queries need not increase as density decreases because the remaining search cost also depends on the basin transitions in Fig.~\ref{fig9.1} and Fig.~\ref{fig9.2}. All simulations use Qiskit's Aer ideal simulator without noise modeling.

\begin{table*}[!t]
\caption{Cumulative query costs and reductions. Aer entries give the mean $\pm$ sample standard deviation (SD) over 300 random starting points. Reductions are baseline minus framework, computed before rounding; all entries are in query-cost units. The DEM reference specifies a density error without selecting a particular estimator and is therefore evaluated analytically only. In Aer, WL supplies density estimates within the adaptive framework, including when its DEM branch is selected.}
\label{tab:cumulative_summary}
\centering
\fontsize{9.5}{11.5}\selectfont
\setlength{\tabcolsep}{3pt}
\renewcommand{\arraystretch}{1.15}
\begin{tabular*}{\textwidth}{@{\extracolsep{\fill}}lcccccccc@{}}
\toprule
& & \multicolumn{4}{c}{Cumulative query cost}
& \multicolumn{3}{c}{Reduction from baseline}\\
\cmidrule(lr){3-6}\cmidrule(l){7-9}
Dataset & $c$ & BBHT & FP & DEM & Framework & BBHT & FP & DEM\\
\midrule
\multicolumn{9}{@{}l}{\textit{Analytical expected queries}}\\[2pt]
Deceptive trap & 0 & 11.5 & 15.4 & 8.4 & 8.0 & 3.5 & 7.3 & 0.4\\
Deceptive trap & 1 & 20.8 & 17.9 & 10.0 & 9.7 & 11.1 & 8.3 & 0.3\\
\addlinespace[2pt]
0/1 knapsack & 0 & 9.5 & 14.0 & 7.5 & 7.4 & 2.2 & 6.6 & 0.2\\
0/1 knapsack & 1 & 19.1 & 27.9 & 9.5 & 9.0 & 10.1 & 18.9 & 0.5\\
\addlinespace[2pt]
\midrule
\multicolumn{9}{@{}l}{\textit{Aer queries: mean $\pm$ sample SD (300 runs)}}\\[2pt]
Deceptive trap & 0 & $10.8\pm13.9$ & $14.3\pm16.0$ & --- & $7.8\pm8.9$ & 3.0 & 6.5 & ---\\
Deceptive trap & 1 & $19.4\pm21.6$ & $16.7\pm18.7$ & --- & $10.1\pm9.5$ & 9.3 & 6.5 & ---\\
\addlinespace[2pt]
0/1 knapsack & 0 & $9.6\pm8.1$ & $12.6\pm14.3$ & --- & $7.8\pm5.2$ & 1.8 & 4.8 & ---\\
0/1 knapsack & 1 & $19.4\pm13.5$ & $25.2\pm28.5$ & --- & $9.2\pm6.3$ & 10.2 & 16.0 & ---\\
\bottomrule
\end{tabular*}
\end{table*}

\begin{figure*}[!t]
\centering
\includegraphics[width=\linewidth]{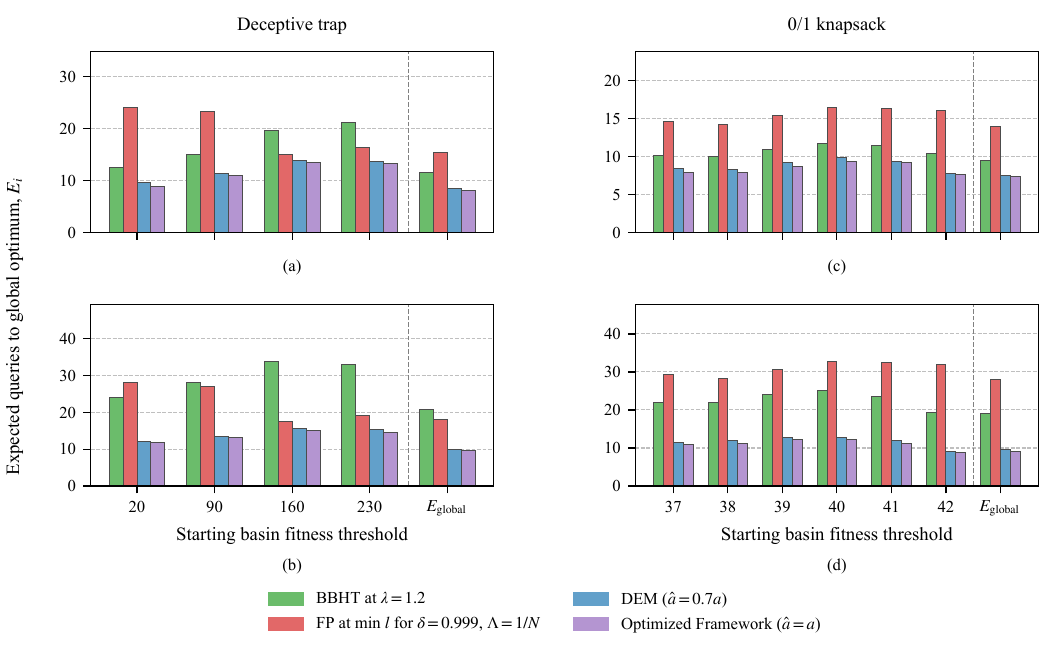}
\caption{Comparisons of analytical queries from certain basins to the global optimum, calculated from \eqref{eq:8.2}. Panels (a)--(b) show the deceptive trap instance and (c)--(d) show the 0/1 knapsack instance, with $c=0$ and $c=1$ in the upper and lower rows, respectively. The rightmost bar ($E_{\mathrm{global}}$) represents the expected average query from all possible points, calculated from \eqref{eq:8.1}.}
\label{fig10.1}
\end{figure*}

Table~\ref{tab:cumulative_summary} summarizes the cumulative queries and their reductions relative to each baseline. The analytical results for the deceptive trap and 0/1 knapsack landscapes are presented in Fig.~\ref{fig10.1}. The analytical optimized framework outperforms the tested BBHT and FP configurations and DEM (at 30\% density underestimation) across both landscapes and cost conventions.

\begin{figure*}[!t]
\centering
\includegraphics[width=\linewidth]{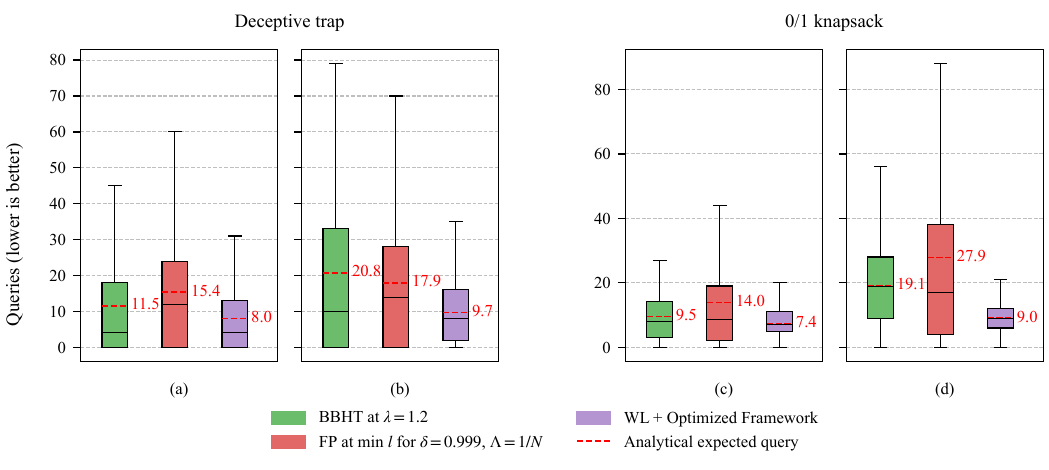}
\caption{Real queries to the global optimum comparisons, using hybrid hill-climbing + GAS on Qiskit's Aer ideal simulator with no noise modeling for 300 uniformly random starting points. Panels (a)--(b) show the deceptive trap instance and (c)--(d) show the 0/1 knapsack instance, with $c=0$ and $c=1$ on the left and right of each pair, respectively.}
\label{fig10.2}
\end{figure*}

As for the actual process execution, real query counts are presented in Fig.~\ref{fig10.2} for the deceptive trap and 0/1 knapsack instances, with $E_{{global}}$ of BBHT, FP, and the optimized framework visualized as dashed lines. The framework has the lowest observed mean cost in all 4 simulation cases. Its mean costs are 7.8 and 10.1 for trap and 7.8 and 9.2 for knapsack, for $c=0$ and $c=1$, respectively. Relative to the tested BBHT and FP configurations, the reductions range from 1.8 to 16.0 query-cost units, or 18.8\%--63.4\% using unrounded means. The larger absolute reductions under $c=1$ include charged preparations as well as oracle queries and should be interpreted under that convention. The smaller analytical advantage over DEM, 0.2--0.5 units, is consistent with the close optimized FP and standard-Grover costs in sparse-density regimes. This advantage is consequently more limited than the reduction against the tested nonoptimized baselines. The reported SD values show substantial run-to-run variability; these comparisons concern observed means, not superiority in every run.

\section{Conclusion and Discussions}
    In the case of Hadamard initialization being free ($c=0$), BBHT is the dominant strategy from 49.9\% down to 3.1\% density on the tested grid; the next point, 3.0\%, favors FP and DEM. Below that range, DEM and both FP cases yield numerically indistinguishable query costs. As for the case where Hadamard initialization is costly ($c=1$), FP with $\Lambda=a$ is the dominant choice at densities above 25\%. 

    When it comes to the practical use-case scenario, this study proposes a potential example framework that combines GAS with Wang-Landau estimation that operates dynamically on the selected method for each estimated density with its corresponding tuned configuration. The experiments are conducted on a 16-variable deceptive trap instance and a 12-variable 0/1 knapsack instance. Theoretical results based on the analytical model demonstrate that, when compared to the 3 tested baselines, including naive setups of BBHT and FP, and DEM with distorted solution-count information, the optimized framework reduces the total queries required to reach the global optimum under both query-counting conventions ($c=0$ and $c=1$). 

    In addition to the analytical results, the real accumulated queries on both test instances are counted in ideal simulations. These data collected from 300 randomly initialized points exhibit results close to the analytical results, as the proposed adaptive framework yields a smaller number of accumulated queries to the optimum point than the tested BBHT and FP configurations, on average.  

    The optimization formulations provided in this study can be directly applied to different density scales, such as smaller density areas for practical implementations. Also, since they are implemented based on direct grid search with limited grid resolution, only the tested configurations are reported. To obtain further precision on these numbers, a deeper level of search is required. 

\bibliographystyle{IEEEtran}
\bibliography{references}

\end{document}